# The delineation of nanoscience and nanotechnology in terms of journals and patents: a most recent update

Loet Leydesdorff
Amsterdam School of Communications Research (ASCoR), University of Amsterdam,
Kloveniersburgwal 48, 1012 CX Amsterdam, The Netherlands.
loet@leydesdorff.net ; http://www.leydesdorff.net

**Abstract**

The journal set which provides a representation of nanoscience and nanotechnology at the interfaces among applied physics, chemistry, and the life sciences is developing rapidly because of the introduction of new journals. The relevant contributions of nations can be expected to change according to the representations of the relevant interfaces among journal sets. In the 2005 set the position of the USA decreased more than in the 2004-set, while the EU-27 gained in terms of its percentage of world share of *citations*. The tag "Y01N" which was newly added to the EU classification system for patents, allows for the visualization of national profiles of nanotechnology in terms of relevant patents and patent classes.

## Introduction

In a recent publication, Leydesdorff & Zhou (2007) delineated (*i*) nanoscience in terms of journals using betweenness centrality in the vector space as an indicator of interdisciplinarity and (*ii*) nano-technology in terms of patents using the newly added tag "Class 977" of the U.S. Patent and Technology Office (USPTO).[1] In the meantime, the European Patent Office (EPO) has made available online the additional tag "Y01N" specifically designed for the nanosciences and nanotechnology (Scheu *et al*., 2006; Hullmann, 2006). Y01N can be decomposed into Y01N2 for Bio-nanotechnology;

[1] In the International Patent Classification (IPC) the field "B82B: Nano-structures: Manufacture and treatment thereof" corresponds to the special class CL/977 which was added to the USPTO.

Y01N4 Nanotechnology for information processing, storage and transmission; Y01N6 Nanotechnology for materials and surface science; Y01N8 Nanotechnology for interacting, sensing, or actuating; and Y01N10 Nanooptics.

In this study, the journal map for nanoscience and nanotechnology is updated using precisely the same technique, but using data from the latest available *Journal Citations Report 2005.* Within the newly delineated domain a publication and citation analysis will be pursued comparing the leading countries (and the EU-27) in terms of percentages of world share. The new tag of the EPO is used to provide another assessment of the relative strengths and weaknesses of the various countries.

## Delineation of the nano-relevant journal set in 2005

In 2005, twelve journals (Table 1) included in the *Science Citation Index* and its *Journal Citations Report* contain the root "nano" in the title, while this number was only six in 2004 and three in 2003 (cf. Braun *et al*., 2007).

| |
|---|
| *Current Nanoscience* |
| *Fullerenes Nanotubes and Carbon Nanostructures* |
| *IEEE Transactions on Nanobioscience* |
| *IEEE Transactions on Nanotechnology* |
| *Journal of Nanoparticle Research* |
| *Journal of Nanoscience and Nanotechnology* |
| *Microfluidics and Nanofluidics* |
| *Microsystem Technologies--Micro- and Nanosystems-Information Storage And Proc* |
| *Nano Letters* |
| *Nanotechnology* |
| *Physica E-Low-Dimensional Systems & Nanostructures* |
| *Synthesis and Reactivity in Inorganic Metal-Organic and Nano-Metal Chemistry* |

**Table 1**: Twelve journals with the root "nano" in their title included in the *Journal Citation Report 2005.*

Using these 12 journals as seed journals, all journals citing or cited by them to the extent of at least one percent of each seed journal's total citations in the cited and citing dimensions, respectively, will be considered as nano-relevant. These were 142 journals in 2005 (against 67 journals in 2004 and 85 journals in 2003). The expansion in 2005 is due to the inclusion of *IEEE Transactions on Nanobioscience*. This journal makes a set of journals in the life sciences relevant to the citation environment of the nano-group.

Among the 142 journals relevant in 2005, 79 are *cited* by one of the seed journals to the level of more than 1% of the total citations of the respective journal (against 38 journals in 2004). These journals can be considered as the knowledge base for the development of an emerging cluster of nano-science journals. Figure 1 shows the map of these 79 citation patterns after normalization, using the cosine as a similarity measure.

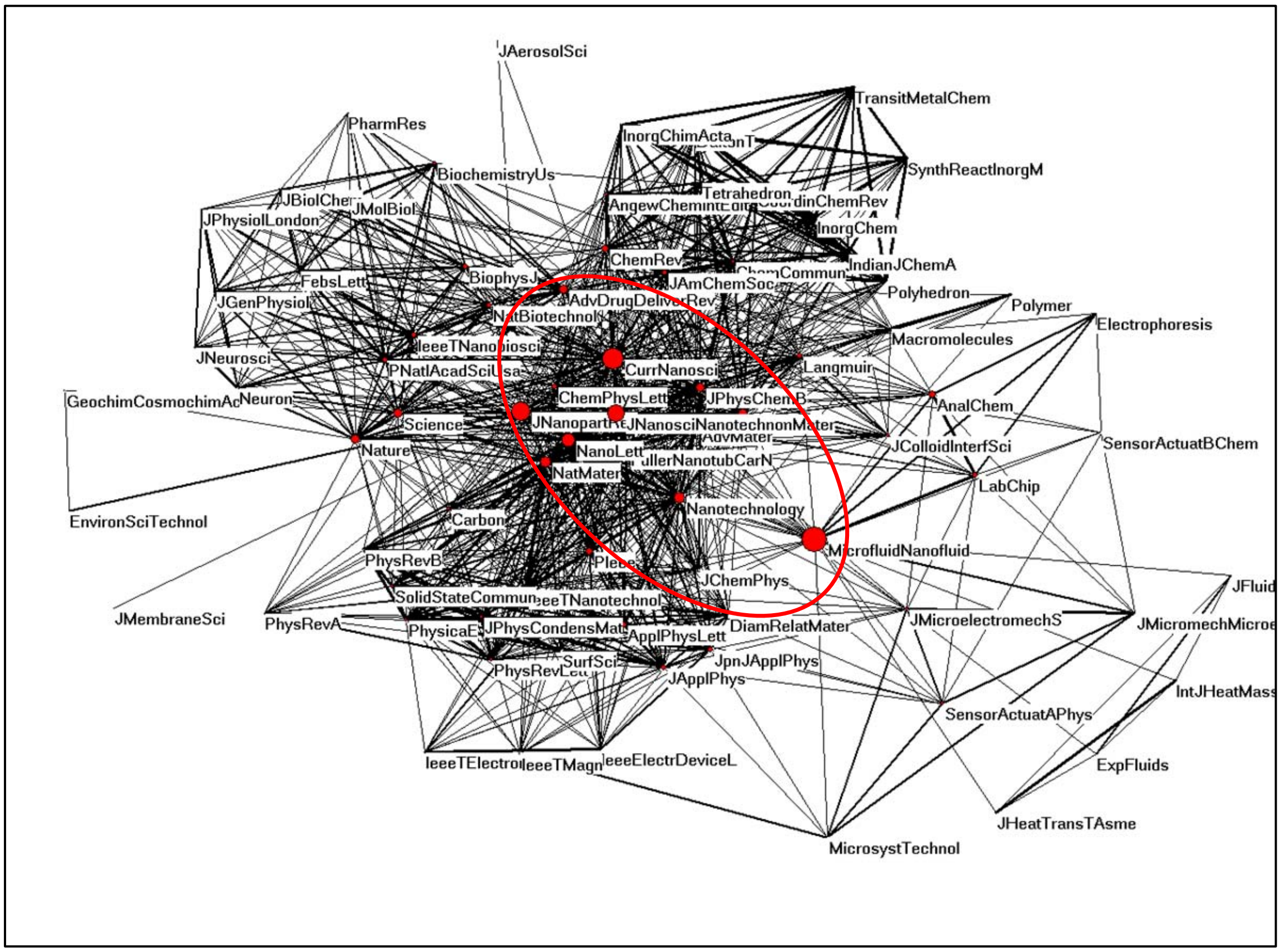

**Figure 1**: Betweenness centrality among the 79 journals cited by 12 nano-journals to the extent of more than one percent of their respective citation totals (cosine ≥ 0.2).

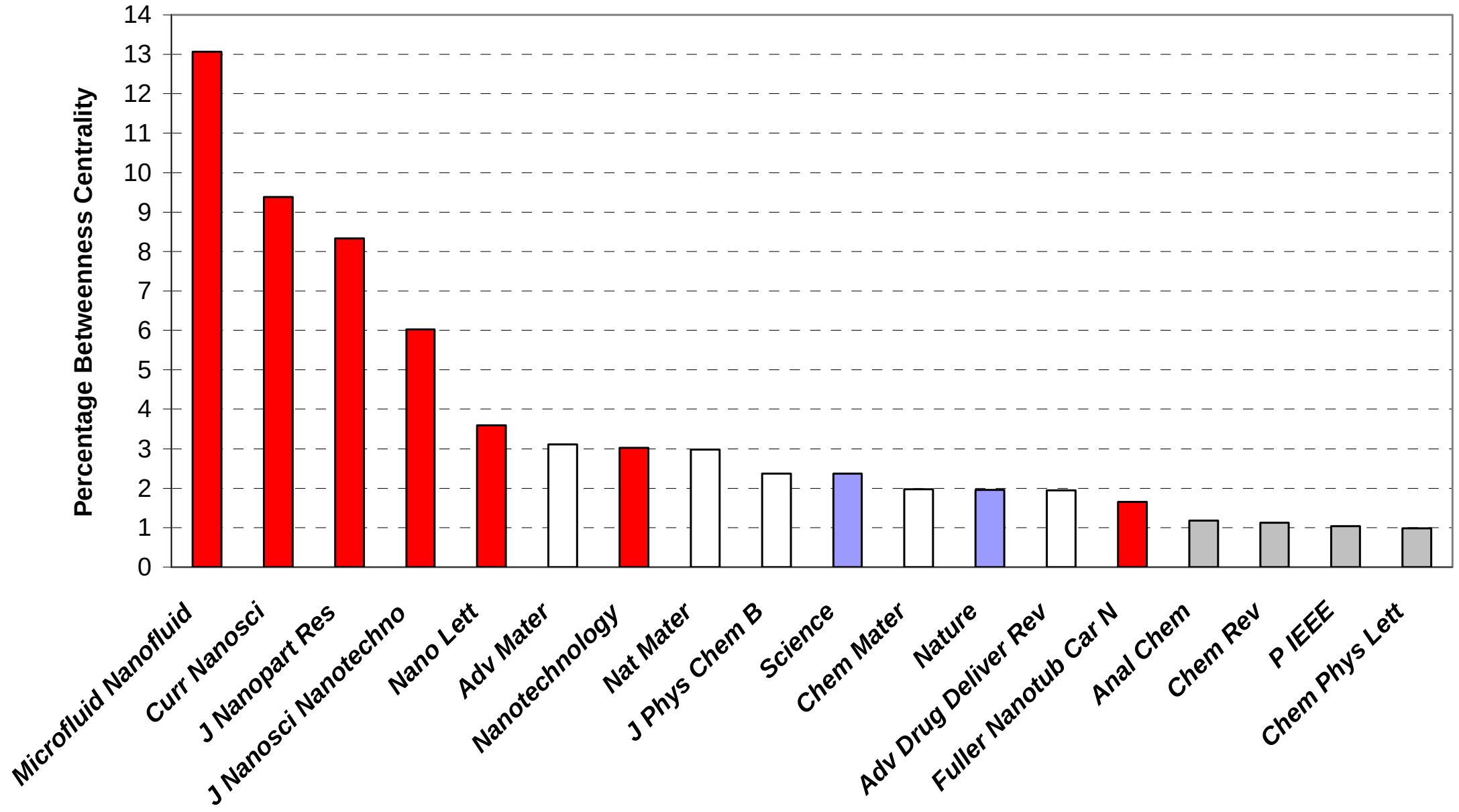

**Figure 2**: Percentage of betweenness centrality among the 79 journals cited by 12 nano-journals to the extent of more than one percent of the citation totals (cosine ≥ 0.2).

Figure 2 shows the journals with a percentage of betweenness centrality larger than one in declining order. The seven journals indicated in red (dark grey) belong to the original set of 12 journals with "nano" in their titles. As expected, *Science* and *Nature* show a high betweenness centrality, but these journals are not specifically part of the nano-set. The five journals indicated in white would belong to the nano-set using this criterion. These twelve journals (Table 2) are encompassed by a red line on the map in Figure 1.

| |
|---|
| *Advanced Drug Delivery Reviews* |
| ***Advanced Materials*** |
| ***Chemistry of Materials*** |
| *Current Nanoscience* |
| ***Fullerenes Nanotubes and Carbon Nanostructures*** |
| ***Journal of Nanoparticle Research*** |
| ***Journal of Nanoscience And Nanotechnology*** |
| ***Journal of Physical Chemistry B*** |
| *Microfluidics and Nanofluidics* |
| ***Nano Letters*** |
| ***Nanotechnology*** |
| *Nature Materials* |

**Table 2**: Twelve journals with betweenness centrality larger than one percent in the relevant citation environment.

I have boldfaced the eight journals in this set which were also included in the corresponding set of ten journals in 2004. *IEEE Transactions on Nanotechnology* and the *Journal of Materials Chemistry* are no longer part of this group, but four new journals are included. *IEEE Transactions on Nanobioscience* is not part of this core group, but a journal firmly embedded in the set of life-science journals identifiable at the top-left corner of Figure 1. *Advanced Drug Delivery Reviews* is less well integrated into this latter group, and seems to contain articles sharing the interdisciplinarity pattern of citations which generates betweenness centrality (Leydesdorff, 2007).

In other words, journals in nanoscience and nanotechnology are not (yet) a stabilized set. The set is developing at interfaces, and the introduction of new journals such as *Nature Materials* and *Current Nanoscience* changes these interfaces. Using other algorithms (e.g., *k*-core), one can show that the entire nano-set has more clearly become a part of applied physics, although the betweenness centrality in the citing dimension indicates the relative dependency of the set on input—in the form of flows of citations—from chemistry and the life sciences.

## Performance within the core set

Because the research front is still so much in flux in terms of the relevant journals, one can expect that the changing journal delineation may have an effect on the relative standing of the various countries. Figure 3 shows that in the new set the USA has a percentage share of publications larger than that of the EU-27, while the EU-25 had a considerable lead in terms of publications using the 2004-set. However, in this set the share of the USA was declining faster than that of the EU. Since percentages of shares are a zero sum game, these declines are related to the strong increase in the percentage of the world share of China. The increase of China is even more pronounced than in the 2004 set (Leydesdorff & Zhou, 2007, at p. 707; cf. Kostoff, 2004, 2007; Zhou & Leydesdorff, 2006).

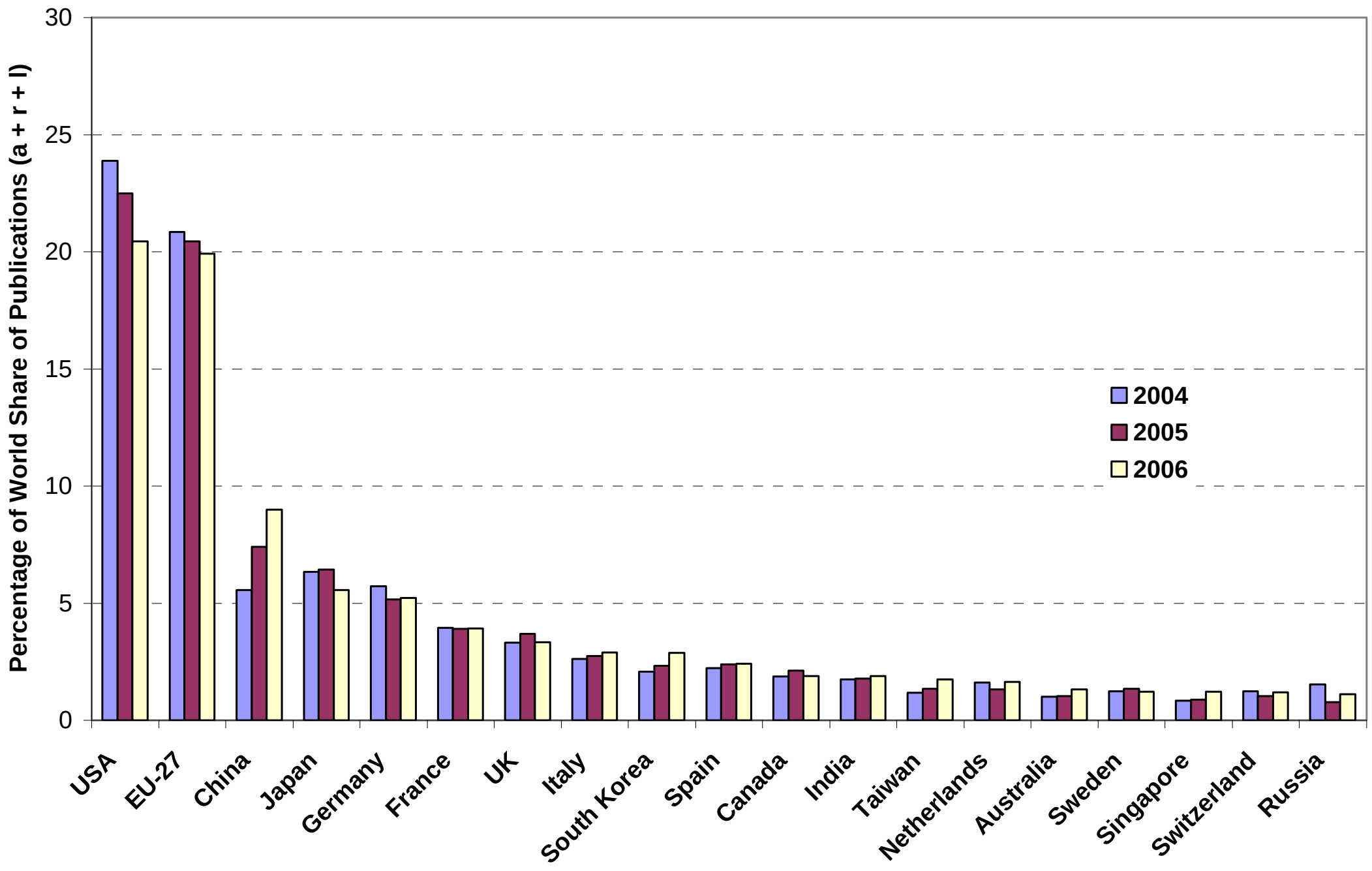

**Figure 3**: Percentage of world share of publications in twelve core journals of nanoscience and nanotechnology (2004-2006; integer counting).

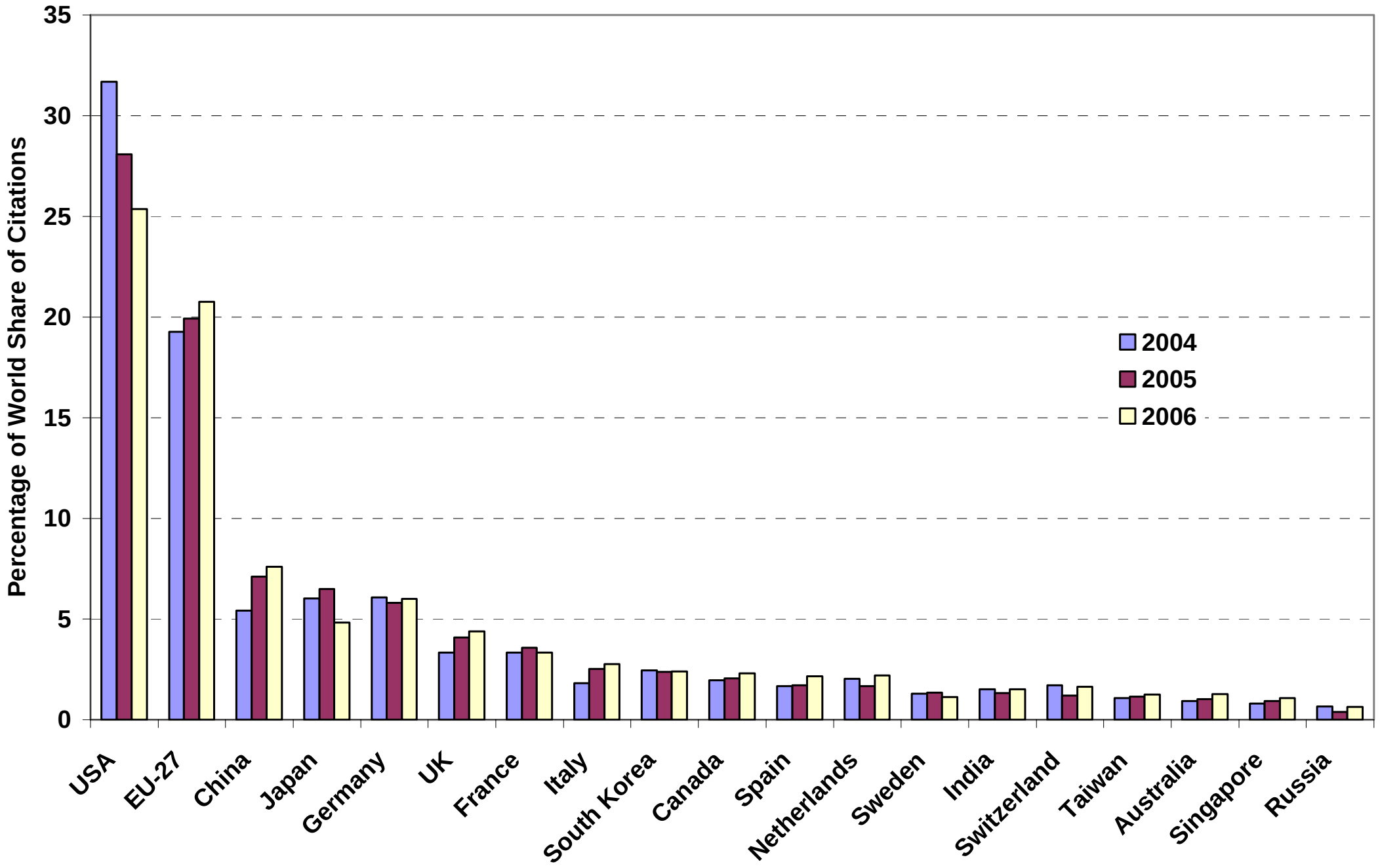

**Figure 4**: Percentage of world share of citations of publications in twelve core journals of nanoscience and nanotechnology (2004-2006; measurement using the *SCI-Expanded* on April 24, 2007).

Figure 4 provides the corresponding figure for the frequency with which documents in these twelve journals have been cited in three consecutive years (as measured on April 24, 2007). Unlike the previous analysis for the 2004 set, the EU-27 no longer loses ground in this set to China, but the USA does. The USA remains by far the strongest player as a single nation (Leydesdorff & Wagner, forthcoming; Sheldon, forthcoming; Sheldon & Holdridge, 2004). Note that South Korea no longer improves its share using this indicator (Leydesdorff & Zhou, 2005).

### The new patent category "Y01N"

In the context of a study about using co-classification as a mapping tool for patents Leydesdorff (2008) added the tag "Y01N" to a dataset which comprises the 135,781 so-called PCT patents published in 2006. (PCT stands for the Patent Cooperation Treaty in which 135 nations subscribe to the possibility of filing patents through the World Intellectual Patent Organization [WIPO] in Geneva.) The files were brought online at http://www.leydesdorff.net/wipo06 for each of the 126 countries with patents in 2006, after proper normalization using the cosine and in the Pajek format.[2] After importing a file into Pajek, the user can select the $k = 1$ neighborhood for the category "Y01N." When this is done for the full set one obtains Figure 5 based on the 762 PCT-patents tagged with Y01N in 2006.

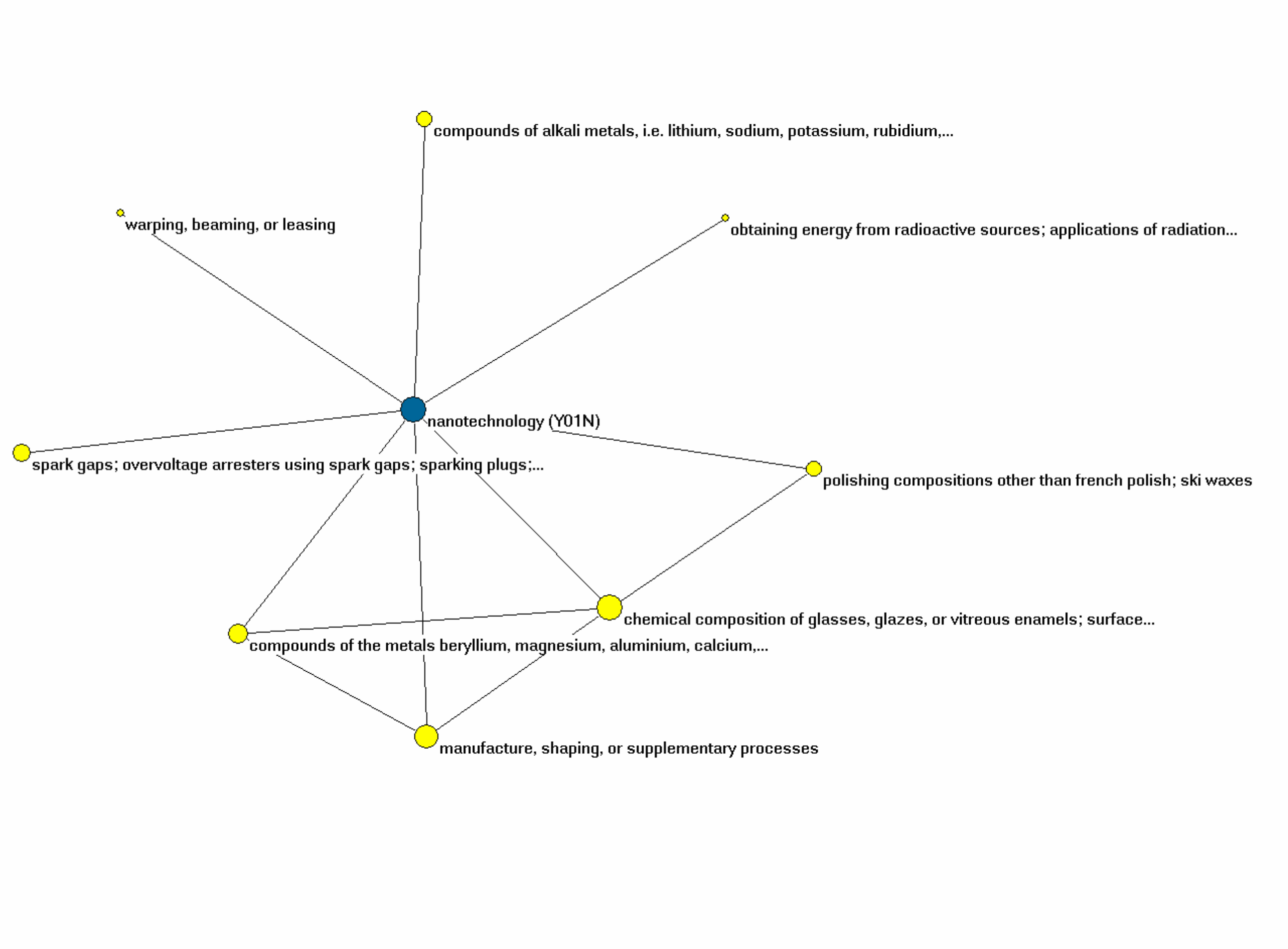

**Figure 5**: $k = 1$ neighborhood of class Y01N; $N = 762$; cosine $\geq 0.05$. (Visualization based on the algorithm of Kamada & Kawai, 1989.)

The distribution of these patents over the countries is as follows:

[2] The freeware program Pajek is available for academic purposes at http://vlado.fmf.uni-lj.si/pub/networks/pajek/

| | | | |
|---|---|---|---|
| USA | 330 | Austria | 4 |
| Japan | 120 | Australia | 4 |
| Germany | 88 | India | 4 |
| France | 46 | Denmark | 3 |
| United Kingdom | 34 | Greece | 3 |
| South Korea | 23 | Norway | 3 |
| Netherlands | 21 | Poland | 3 |
| Switzerland | 15 | Russia | 3 |
| Italy | 15 | Brazil | 2 |
| Canada | 13 | New Zealand | 2 |
| China | 11 | Turkey | 2 |
| Israel | 7 | Belarus | 1 |
| Sweden | 7 | Czech Republic | 1 |
| Belgium | 6 | Hong Kong | 1 |
| Spain | 6 | FYR Macedonia | 1 |
| Singapore | 6 | Mexico | 1 |
| Finland | 5 | Romania | 1 |
| Ireand | 5 | Taiwan | 1 |
| | | South Africa | 1 |

**Table 3**: The distribution of patents over (37) countries for the category "nanotechnology" (Y01N) using the WIPO dataset 2006 (762 patents; 799 addresses).

Using the online files for specific countries one is able to draw similar pictures, but country-specific. Figure 6 provides as an example a visualization of the co-classification of patents with a French address among the inventors. Although based on only 46 patents these figures are more informative than the overall set. Aggregation averages out the nation-specific profiles.

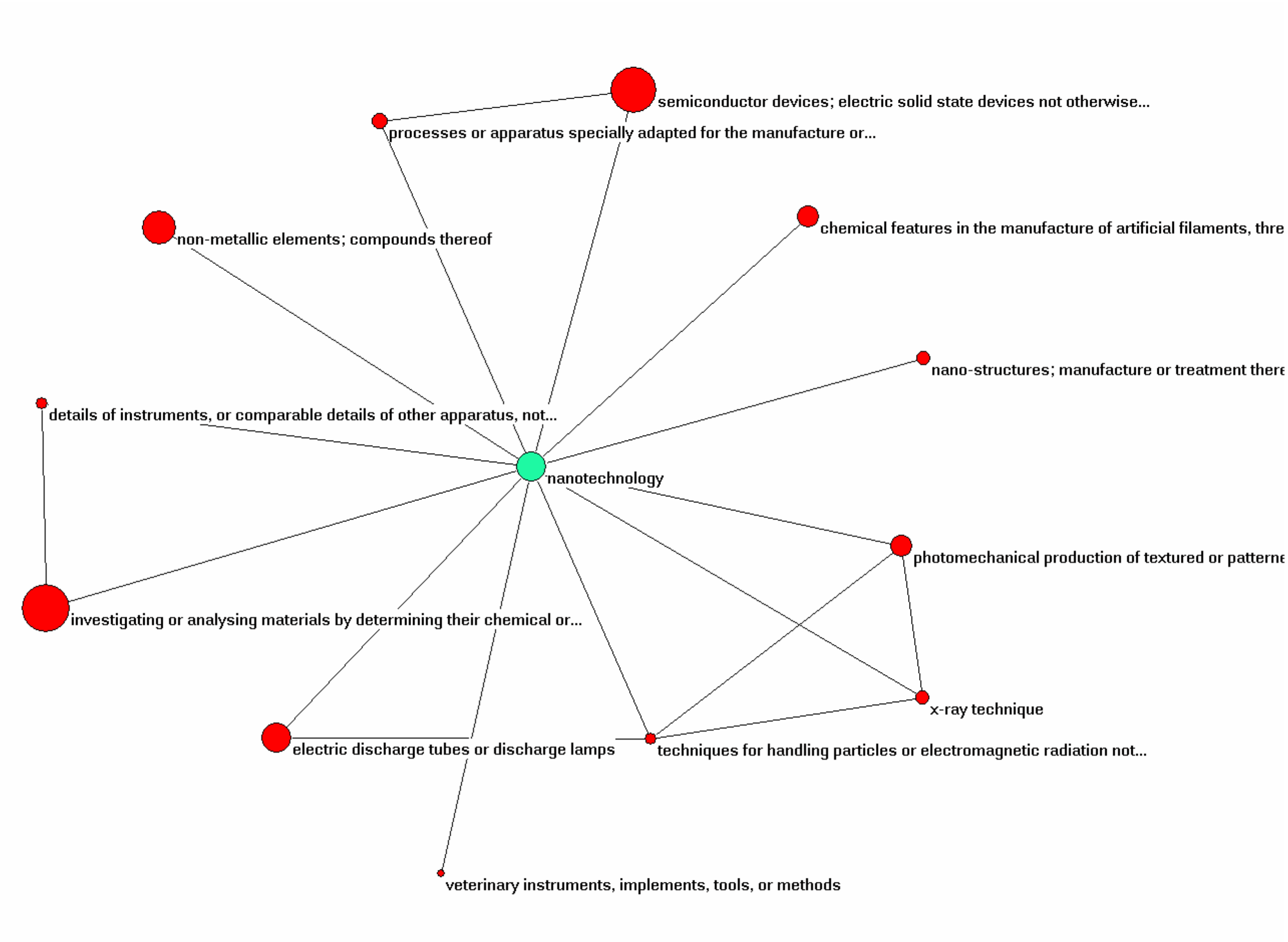

**Figure 6**: *k* = 1 neighborhood of class Y01N for 46 patents having at least one inventor with a French address; cosine ≥ 0.05.

## Conclusions

- Using betweenness centrality in the vector space, a set of 12 journals can be indicated as developing interdisciplinarily at the interfaces between applied physics, chemistry, and the life sciences.

- The focus of activity in the thus delineated journal set shifted from 2004 to 2005. The position of the EU hass improved in the new set compared with the USA in terms of citations, but not in terms of publications. These two major players and Japan are further losing ground to China, which is not only increasing its percentage of world share of publications but also its percentage of world share of citations in this set, albeit the latter at a lower rate.

- The new tag of nanotechnology developed by the European Patent Office can now be used for retrieval and analysis. This tag allows for more detailed analysis in subcategories like nano-optics and bio-nanotechnology. Country-specific maps exhibit national profiles in nanotechnology.